# Quantum Kernel Anomaly Detection Using AR-Derived Features from Non-Contact Acoustic Monitoring for Smart Manufacturing

Takao Tomono
*Graduate School of Science and Technology*
*Keio University*
Kanagawa, JAPAN
takao.tomono@ieee.org

Kazuya Tsujimura
*Digital Innovation Divison*
*TOPPAN Holdings Inc.*
Tokyo, JAPAN
kazuya.tsujimura@toppan.co.jp

*Abstract*—The evolution of manufacturing toward Smart Factories has highlighted critical challenges in equipment maintenance, particularly the reliance on numerous contact sensors for anomaly detection, resulting in escalating sensor and computational costs. This study investigates the application of quantum kernels to enhance anomaly detection using non-contact sensors. We hypothesized that quantum computing's expressive power could effectively discriminate among multiple anomaly types using fewer sensors. Our experimental setup involved detecting and classifying anomalies from two distinct manufacturing equipment: a conveyor and a chain belt machine using a single directional microphone positioned at varying distances (0-3m). Audio data was processed through Autoregressive (AR) models to extract coefficient features, which were then mapped into quantum feature space using quantum kernels for one-class SVM classification. Results demonstrated that quantum kernel implementations maintained near-perfect accuracy and F1-scores (consistently >0.92) across all distances, while classical approaches showed significant performance degradation beyond the 0m position. Feature space visualization revealed that quantum kernels effectively separated different anomaly types into distinct quadrants within a two-dimensional representation, enabling not only detection but also classification of multiple equipment failures. Specifically, under the third and fourth features space, conveyor anomalies consistently appeared in the second quadrant, while chain belt anomalies clustered in the fourth quadrant. This study demonstrates that quantum kernel methods enable significant anomaly detection in noisy factory environments using fewer non-contact sensors, representing an important step toward realizing quantum-enhanced smart factories with reduced infrastructure requirements and improved maintenance efficiency.

## I. Introduction

The proliferation of IoT devices in manufacturing has generated vast datasets that can address critical maintenance challenges. However, conventional anomaly detection approaches rely heavily on numerous contact sensors, resulting in escalating costs and complexity as production systems scale [1], [2], [3].

Grant [No.] : NEDO[JPNP23003], JST[JPMJPF2221]

Effective maintenance of production equipment is fundamental to manufacturing efficiency and product quality. Conventional approaches to equipment maintenance rely on constructing learning models using data acquired from multiple vibration sensors physically attached to devices [4], [5], [6], [7], [8], [9]. As the number of devices increases, the number of sensors grows exponentially, resulting in significant time requirements for developing classification models for individual anomalies. This leads to a dramatic increase in both sensor and computational costs.

Traditional neural network-based learning models typically require thousands to tens of thousands of data points. However, production environments with high-mix, low-volume manufacturing are not well-suited for building learning models that require extensive datasets. Even after implementing anomaly detection models in factories, identifying which equipment or component is experiencing an abnormality can be time-consuming. In practice, production site personnel often walk around the factory, directly listening for anomalies, which heavily relies on individual experience and intuition.

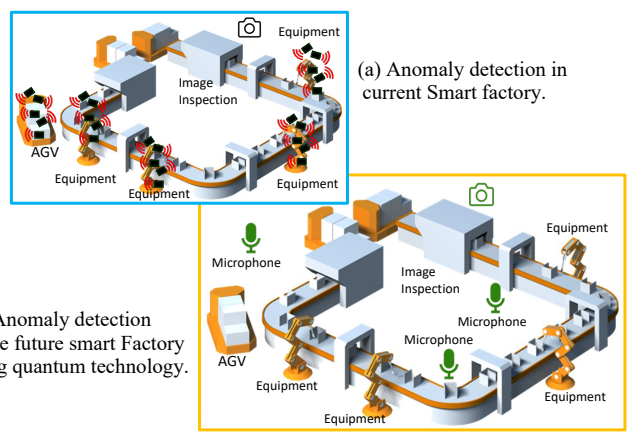

Fig. 1. Anomaly detection approaches: (a) Current contact sensor-based systems; (b) Proposed quantum-enhanced non-contact approach.

As shown in Fig.1, we anticipate the evolution from conventional ICT-based Smart Factories to future Smart Factories utilizing quantum technologies. The figure illustrates that multiple vibration sensors are installed on each manufacturing device. Many contact-type sensors are required to detect anomalies based on vibration frequencies generated by motors and the inherent resonance frequencies of equipment. In the future, we expect that a smaller number of non-contact sensors, such as microphones, will be able to detect multiple anomalies across numerous production facilities. The establishment of such anomaly detection systems is expected to contribute to reducing sensor wiring and power consumption.

Generally, time series data can be modeled using AR, MA, ARMA, ARIMA, SARIMA, and other methods, with the model selection depending on the data characteristics[10], [11], [12], [13], [14]. While these models are used for anomaly detection, Support Vector Machine (SVM) is frequently employed for classification tasks. SVMs demonstrate high discrimination capabilities among various machine learning techniques and are widely used.

Particularly, kernel-based SVMs are extensively utilized in pattern recognition because they can effectively separate nonlinear spaces. However, kernel-estimated SVMs encounter limitations as the computational time increases significantly when feature spaces become large. Conversely, quantum approaches attempt to improve computational speed by representing large feature spaces. Havlíček et al. proposed a quantum solution by introducing quantum entanglement and Pauli Z feature maps into exponentially large feature spaces [15]. By using a new design called the projected quantum kernel, H-Y. Huang et al. have succeeded in demonstrating large-scale quantum advantages that could not be achieved using existing methods [16].

We are focusing on the potential expressive power of quantum kernel method [17], [18], [19]. We hypothesize that quantum kernels can construct more complex nonlinear decision boundaries in AR-derived feature space, enabling classification of multiple anomaly types with higher accuracy than classical methods. Based on this hypothesis, we aim to develop a system that can detect multiple anomalies using several non-contact sensors.

Our approach consists of three stages: (1) acoustic signal acquisition using directional microphones, (2) autoregressive (AR) feature extraction, and (3) quantum kernel-based anomaly classification applied to the AR coefficients. This methodology applies quantum kernels to structured AR features rather than raw time series data, improving classification accuracy in noisy industrial environments. This approach yields three contributions: (1) quantum kernel methods effectively detect and classify multiple equipment anomalies using a single non-contact sensor; (2) quantum implementations maintain high performance at distances where classical approaches fail, enabling flexible sensor placement; and (3) we provide a visual framework that maps anomaly types to specific feature space regions for intuitive diagnosis.

## II. CREATION OF DATASETS

As shown in Table 1, we constructed a comprehensive dataset comprising 60 normal/normal (0/0), 30 anomaly/anomaly (1/1), 30 anomaly/normal (1/0), and 30 normal/anomaly (0/1) samples, where "0" denotes normal operation and "1" indicates anomalous conditions. Of these, 40 normal/normal (CON/CHA=0/0) samples were used as training data, with the remainder serving as test data.

Fig. 2 illustrates the experimental setup for dataset creation. As shown in Fig. 2(a), we utilized two distinct devices—a conveyor (CON) and a chain belt machine (CHA). To simulate anomalous conditions, we strategically inserted nails into both devices to generate characteristic abnormal sounds that represent common failure modes. Meanwhile, sound from the rubber belt machine served as environmental noise (white noise). The directional sound collector had microphones installed at 45-degree intervals in a 360-degree direction.

TABLE I. DATASETS COMPOSITION FOR TRAINING AND TESTING

| CON/CHA | 0 / 0 | 0 / 1 | 1 / 0 | 1 / 1 |
|---|---|---|---|---|
| Total | 60 | 30 | 30 | 30 |
| Training | 40 | --- | --- | --- |
| Testing | 20 | 30 | 30 | 30 |

Here, normal present for 0 and Anomaly present for 1.

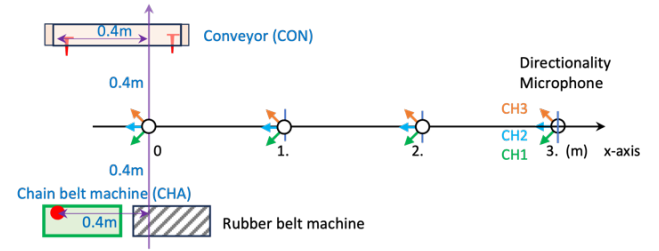

(a) Side View on the measurement on Normal and Anomaly

| | 0m | 1m | 2m | 3m |
|---|---|---|---|---|
| CH3 | CON | CON+CHA | CON+CHA | CON+CHA |
| CH2 | ( CON+CHA ) | CON+CHA | CON+CHA | CON+CHA |
| CH1 | CHA | CON+CHA | CON+CHA | CON+CHA |

(b) The equipment sound of detection by each channel.

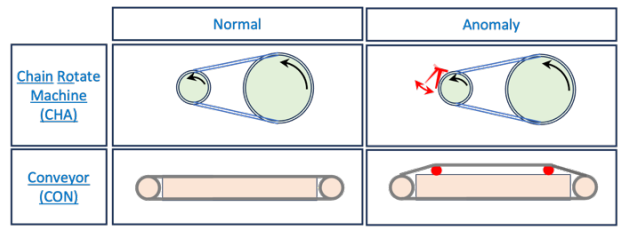

(c) Plain View on the measurement on normal and anomaly sound

Fig. 2. Experimental setup: (a) Equipment layout and measurement positions; (b) Channel detection capabilities; (c) Normal vs. anomalous conditions.

We use three channels (CH1, CH2 and CH3). The directional microphone sound was positioned at distances of 0m, 1m, 2m, and 3m, recording both normal and anomaly sounds at each position. Fig.2 (b) shows which devices could be recorded by CH1,CH2 and CH3. Fig. 2 (c) illustrates anomaly data was created using nails. For CHA anomaly (1), the nail tip was positioned to hit the chain belt in the rotation direction, causing the nail to swing pendulum-like, alternately hitting and separating from the belt. Normally, a rubber belt runs on a conveyor while touching a metal plate. However, by clamping two nails between the rubber belt, the rubber belt runs partly in the air above the metal plate.

### III. QUANTUM KERNEL

Quantum kernel method represent a promising approach to leveraging quantum computing's capabilities within the current NISQ (Noisy Intermediate-Scale Quantum) era. While classical kernels map data into higher-dimensional feature spaces to improve separability, quantum kernel approach utilize quantum state spaces that can be exponentially larger than classical counterparts. The quantum kernel function is defined as:

$$\kappa(x_i, x_j) = \left|\left\langle \phi(x_j)^\dagger \middle| \phi(x_i) \right\rangle\right|^2 \quad\quad 1$$

Where $\phi(x_j)$ represents a quantum feature map that encodes classical data points $x_i$ into quantum states $|\phi(x_j)\rangle$. This inner product represents the quantum state overlap and serves as a similarity measure between data points. The feature map typically employs parameterized quantum circuits with operations that create entanglement, thereby accessing feature spaces that would require exponentially many dimensions classically.

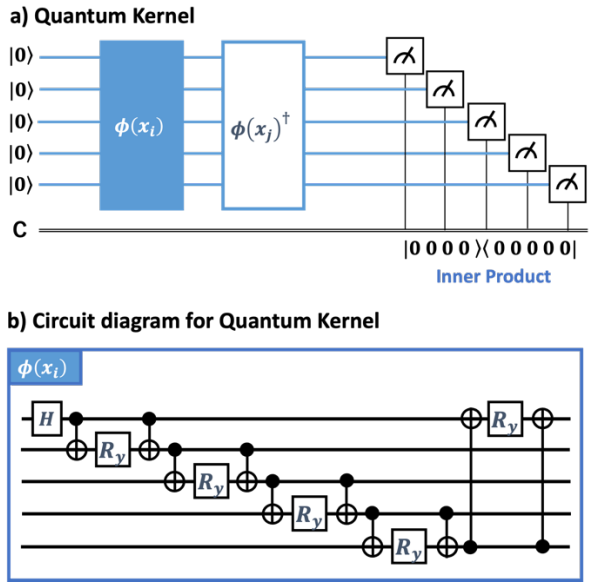

Fig.3. Quantum kernel implementation details on 5 qubits: (a) Quantum kernel circuits diagram. (b) Detailed quantum circuit diagram implementing with encoding rotations Ry (Y-feature map) and entangling operations.

Havlíček et al. [15] introduced a framework for supervised learning using quantum-enhanced feature spaces, demonstrating that quantum kernel approach could potentially offer advantages for certain classification problems. Recent theoretical work by Liu et al. [23] established rigorous conditions under which quantum kernel approach can provide provable computational advantages over classical approaches. Additionally, Huang et al. [16] explored how the power of quantum kernel scales with dataset size, showing potential advantages in the small data regime—a characteristic particularly valuable for industrial settings where anomaly data is scarce.

In our implementation, we utilize a quantum feature map with entangling operations (Fig. 3) that projects our 5-dimensional AR coefficient vectors into a much higher-dimensional quantum state space. This approach potentially enables more complex decision boundaries in the SVM classifier, enhancing the separability of anomaly patterns even when signal quality is degraded by distance and environmental noise.

### IV. PROPOSED METHOD AND ANALYSIS

Most products and manufacturing equipment are designed to minimize failures. In such cases, unsupervised learning is commonly used for anomaly detection. In this study, we extracted coefficient parameters $\phi_i$ (i: feature value) from the autoregressive (AR) model of vibration data (time series data) as sound. These features were used in one-class SVM to determine anomalies[20]. Fig.4 illustrates the flow from dataset to learning model construction and discrimination. The stored data is fitted to the AR model, and features are extracted using the coefficient parameters. The data is then divided into a training set and a test set. Training is performed using an AR model using normal data. The AR(*p*) model is defined as:

$$\mathcal{Y}_t = c + \sum_{i=1}^{p} \phi_i \mathcal{Y}_{t-i} + \epsilon_t, \epsilon_t \sim W.N.(\sigma^2) \quad\quad 2$$

Where $\epsilon_t$ represents white noise. The environmental sounds of the room and the sound of the rubber belt machine were used as white noise.

The AR(*p*) model parameters are estimated using the Yule-Walker equations [21] and the Levinson-Durbin recursion algorithm [22], which efficiently compute the optimal coefficients by minimizing the prediction error. We selected *p*=5 based on preliminary experiments examining the Akaike Information Criterion (AIC) and Bayesian Information Criterion (BIC) [23] across different model orders, finding that *p*=5 provides the optimal balance between model complexity and goodness of fit for our audio data.

For anomaly detection, we implemented one-class SVM with two kernel variants: (1) a classical Radial Basis Function (RBF) kernel with the standard formulation $\kappa(x_i, x_j) = exp.(-\gamma ||x - y||^2)$, where $\gamma$ was optimized through cross-validation; and (2) a quantum kernel implemented using a Y-feature map ($R_y$) and entangle operation as illustrated in Fig. 3. The quantum circuits were simulated using Qiskit (version 0.42.0) with a state-vector simulator backend.

The one-class SVM was trained exclusively on normal operating data (CON/CHA = 0/0) to learn the boundary of normal behavior in feature space. During testing, samples falling outside this boundary are classified as anomalies. We employed a standard one-class SVM formulation, which controls the upper bound on the fraction of training errors and the lower bound on the fraction of support vectors.

The learning model constructed by training results were tested by using test data. Testing looked at the performance metrics of accuracy and F1 score. Testing was performed under the conditions of CON/CHA=0/0, 0/1, 1/0, and 1/1. Testing was performed both overall across all cases and on individual cases.

As a means of discrimination, the data are plotted in the third and fourth feature spaces.

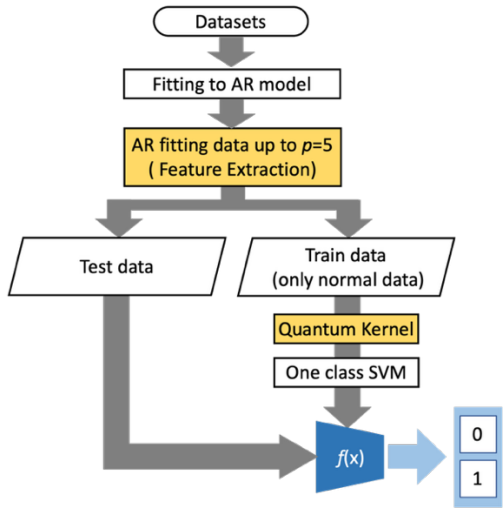

Fig.4. The flow from dataset to learning model construction and discrimination.

## V. RESULTS AND DISCUSSION

### A. Environmental Sound and Individual Sound

Figure 5 presents sound pressure level analysis conducted across five independent measurement series (3000 points each) in a multi-device environment designed to simulate realistic factory acoustic conditions. The experimental setup included two background noise sources: a rubber belt machine (White Noise 1) providing consistent industrial background noise, and ambient room noise measured at four corners (White Noise 2) to establish baseline environmental levels.

The relative positions of individual machines are shown in Fig.2 (a). We are thinking that this situation closely resembles actual factory settings where multiple sound sources compete. The measurements are the average of a series of five data sets.

In Fig.5 (a), we analyze the acoustic profile of individual equipment operation. The CON exhibited sound pressure levels of 43.6-43.8 dB when measured at 0 m distance. However, as the measurement distance increased, the sound pressure levels decreased substantially, reaching 39.7-40.3 dB at the 2 m and 3 m measurement points. Meanwhile, the CHA showed measurements value of 40.4-42 dB at 0 m. And as distance increases, the sound pressure level declines, reaching 36.4-36.8 dB at the 2 m and 3 m. The sound pressure ratio was consistently higher for CON than for CHA, regardless of distance. For the CHA, the difference between normal and anomaly sounds was clearly distinguished at the 0 m and 1 m point. At the remaining measurement locations, no significant difference was observed. Furthermore, at 2 m and 3 m, the sound pressure of the CHA was close to that of white noise1 and 2.

Conversely, Fig. 5 (b) examines cases where two sound pressures exist under white noise 1 and 2. Here, we compare the intensity level of CON/CHA = 0/0, 0/1, 1/0, and 1/1 at 0 m, 1 m, 2 m, and 3 m. The intensity level decreases as distance increases. When multiple equipment operate, the combined sound pressure of CON and CHA would exceed 80 dB. However, the actual measurement was approximately 46 dB. This indicates that the sound pressure level was not the sum of each equipment's sound pressure, but rather an increase of 4-5 dB from the sound pressure of a single equipment.

This phenomenon is likely attributable to interference occurring from various overlapping sounds. Additionally, it demonstrates that the directional microphone effectively detects sound pressure in the direction it is oriented. Based on these findings, we will conduct further analysis using this equipment and microphone positioning.

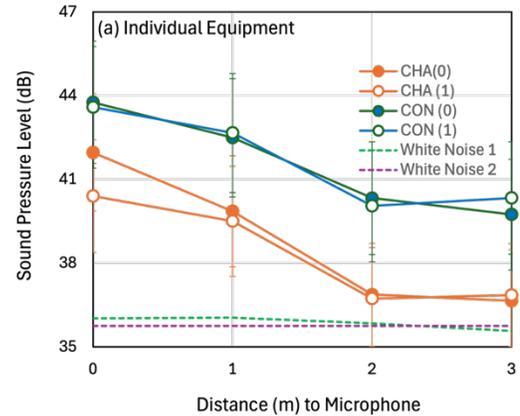

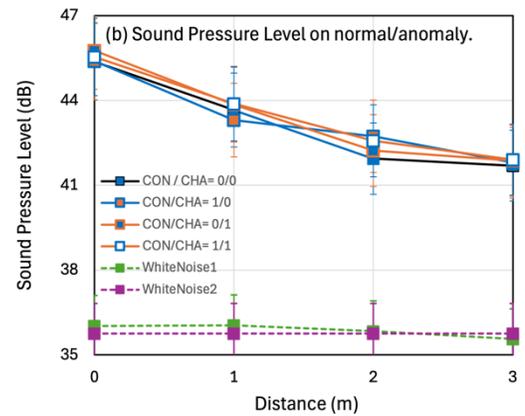

Fig. 5. Sound pressure level analysis: (a) Individual equipment and background noise profiles showing distance-dependent attenuation; (b) Sound Pressure Levels on normal/anomaly (0/0, 1/0, 0/1, 1/1). Error bars represent standard error (n=5 trials).

## B. Performance Metrics Across Sensor Distances

Fig.6 illustrates the relationship between performance metrics (accuracy and F1-score) and microphone position, with black lines representing classical kernel performance and orange lines showing quantum kernel approach results. The performance disparity between these approaches becomes particularly pronounced as sensor distance increases.

Fig.6 (a) shows quantum kernel implementations maintained high accuracy and F1-scores. A minor fluctuation was observed at the 1m position, where both metrics showed a slight decline to approximately 0.92 (F1-score) and 0.95(Accuracy). We obtained optimal performance (1.0) at 0m and 3m distances. Due to limited data points, we report individual measurements rather than aggregate statistics.

We think that minor degradation at 1m and 2m positions is due to acoustic interference patterns. This exceptional distance-independent performance can be attributed to the quantum kernel's ability to effectively map the data into a higher-dimensional feature space, creating more sophisticated decision boundaries that remain robust against noise. The quantum kernel approach appears to exploit the inherent structure that transcend the capabilities of classical approaches, particularly robust in the difference of level.

In contrast, Fig. 6 (b) illustrates the performance of the conventional RBF kernel drops significantly with distance. It drops sharply at 2m, but then rises at 3m. Initially, all channels CH1, CH2, and CH3 showed perfect accuracy and F1 score of 1.0 at 0 m, but as the measurement distance increased, the performance decreased. At 2m, the accuracy was 0.46, 0.27, and 0.41, and the F1 score dropped sharply to 0.46, 0.22, and 0.42, but at 3m, the accuracy was about 0.72 and 0.73 for CH1, CH2, and CH3, and the corresponding F1 score also rose to 0.69. This is thought to be due to sound interference, including reflected sound. Overall, only the accuracy and F1 score decreased with distance.

The initial performance at 0m can be explained by the dominance of a single sound source at this distance: CH1 picks up mainly CHA tones, CH3 mainly CON signals, and CH2 seems to pick up both CHA+CON. As the distance increases, each channel picks up a complex mix of CHAN and CON tones, as well as white noise.

There are several theoretical factors that explain why quantum kernels are robust to noise and distance-induced signal degradation. 1) Exponential feature space dimensionality: The quantum feature map projects 5-dimensional AR coefficients into a $2^5 = 32$ -dimensional quantum Hilbert space, which exponentially increases the freedom of decision boundary construction compared to the conventional RBF kernel. 2) Entanglement-enhanced separability: The entanglement operation in this quantum circuit may create correlations between conventionally inaccessible features, enabling the identification of fine patterns that are discernible even under noise. 3) Noise-resistant quantum states: Quantum superposition allows multiple classical states to be encoded simultaneously, which may provide inherent redundancy that is lacking in classical methods. However, these factors remain to be clarified through further experiments.

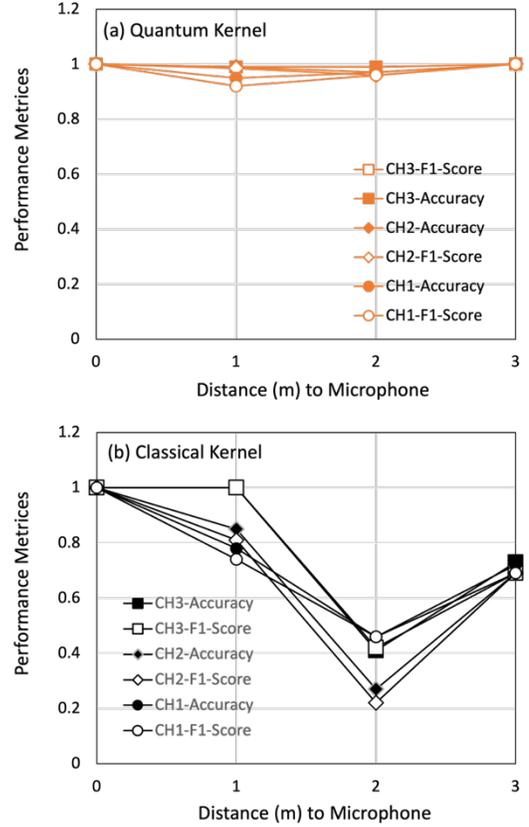

Fig. 6. Performance comparison across measurement distances: (a) Quantum kernel maintains high accuracy and F1-scores (>0.96) at all distances; (b) Classical RBF shows significant degradation beyond 0m. See Table 2 for detailed metrics. The features is 5.

TABLE II. DETAILED PERFORMANCE METRICS ACROSS MEASUREMENT DISTANCES

| Method | metric | 0m | 1m | 2m | 3m | Mean | SD |
|---|---|---|---|---|---|---|---|
| Quantum Kernel | Accuracy | 1.000 | 0.977 | 0.977 | 1.000 | 0.988 | 0.013 |
| | F1-score | 1.000 | 0.965 | 0.970 | 1.000 | 0.984 | 0.018 |
| Classical RBF | Accuracy | 1.000 | 0.877 | 0.380 | 0.713 | 0.743 | 0.257 |
| | F1-score | 1.000 | 0.850 | 0.367 | 0.690 | 0.727 | 0.259 |

Note: Values represent averages across three channels (CH1, CH2, CH3).

Table 2 shows the results of the statistical analysis of the performance differences. For each distance measurement, the performance metrics of the three channels (CH1-3) were averaged. A paired t-test found statistically significant differences between the quantum and classical approaches in terms of means and standard deviations.

- Accuracy: t(3) = 4.46, p = 0.021, Cohen's d = 1.49

- F1 score: t(3) = 4.81, p = 0.017, Cohen's d = 1.61

Effect size analysis using Cohen's d [24] showed a very large practical difference (d > 1.4), significantly exceeding the threshold for a large effect (d = 0.8) established by Cohen (1988). These values indicate that the quantum kernel performs about 1.5 standard deviations better than the classical approach, indicating a large real-world impact beyond mere statistical significance. A large effect size translates into significant operational benefits. The quantum kernel maintains over 96% performance at all distances, while the classical approach drops to 38% at 2m. However, these effect sizes should be interpreted with caution given the limited sample size (n=4 distances) and controlled experimental conditions.

## C. Classification Performance and Error Analysis

Fig.7 presents the confusion matrix for the classical RBF kernel, highlighting a concerning pattern of misclassifications. With 30-34 false Positives recorded. This misclassification result in significant failure rate that would be unacceptable in production environments. where equipment failure can lead to costly downtime or safety hazards.

As reported already[17], In practical terms, quantum approaches may occasionally classify normal conditions as anomalous (warranting further investigation) in initial learning process, but they rarely miss actual anomalies—a characteristic highly desirable in industrial monitoring systems. Conversely, the classical RBF approach demonstrated in this study showed low precision values, resulting in numerous missed anomalies.

For industrial applications, particularly in manufacturing contexts where equipment failures represent significant operational and safety risks, false positives pose a substantially greater concern than false negatives. A false negative may result in a good product failing the test, but potentially passing it on a second inspection. A false positive, on the other hand, may allow equipment failure to progress undetected, leading to catastrophic failures, production line shutdowns, and significantly increased repair costs.

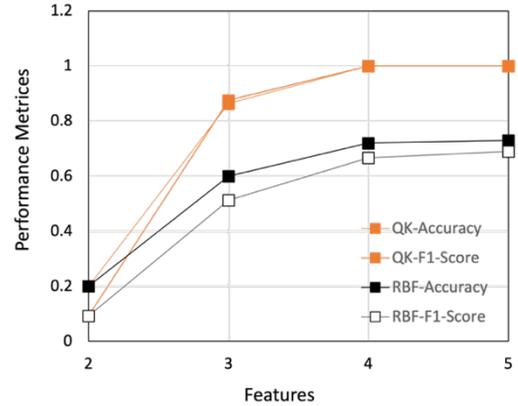

Fig. 7. Confusion matrices for anomaly classification using one-class SVM: (a) Channel CH1 results showing the distribution of true positives (TP), false positives (FP), true negatives (TN), and false negatives (FN) for the classical RBF kernel implementation.

## D. Importance of Features

Fig.8 illustrates the relationship between individual features and overall model performance indicators. Our investigation utilized 5 features (i=5 for coefficient parameters $\phi_i$) for both classical and quantum kernel implementations.

It is generally known that increasing the number of features and increasing the cumulative contribution rate leads to higher performance. Performance metric analysis shows that features 3 and 4 have high contribution rates, with the cumulative contribution rate exceeding 80%. When we used the quantum kernel, the accuracy and F1 score essentially reached a constant states after including the first five features. This means that when we used the quantum kernel, the cumulative contribution rate reaches 100% when the number of features is 5.

Fig.8 The relationship between Features and performance metrics on quantum and classical kernel. The graph illustrates how model accuracy, and F1-scores improve with the sequential addition of AR coefficient features (Feature 1-5).

As mentioned above, we found that features 3 and 4 have a large contribution. The quantum kernel can construct significantly more complex separation surfaces than the classical kernel. Therefore, we hypothesized that a cross-sectional analysis using only the two-dimensional plane formed by features 3 and 4 may reveal a pattern sufficient for distinguishing anomalies.

## E. Two Dimensional Features Space

Fig.9 provides a visualization of anomaly points in the two-dimensional feature space defined by features 3 and 4 when using the quantum kernel. This representation offered remarkable insights into the spatial distribution of different anomaly types.

For channel CH3, when both the conveyor and the chain belt machine were operating normally (CON/CHA), there were few anomalies in the feature space, establishing a clear baseline of normal operation. When the conveyor operated normally but the chain belt exhibited anomalous behavior (CON/CHA=0/1), anomaly points consistently clustered in the fourth quadrant of the feature space. Conversely, when the conveyor showed abnormal operation while the chain belt functioned normally (CON/CHA= 1/0), anomaly points predominantly appeared in the second quadrant.

Most significantly, when both machines operated abnormally (CON/CHA= 1/1), we observed anomaly points distributed across both the second and fourth quadrants, effectively representing the superposition of the individual anomaly patterns. This quadrant-specific distribution provides a powerful visual diagnostic tool that not only detects the presence of an anomaly but clearly identifies which equipment is malfunctioning—a capability that significantly enhances the practical utility of the system.

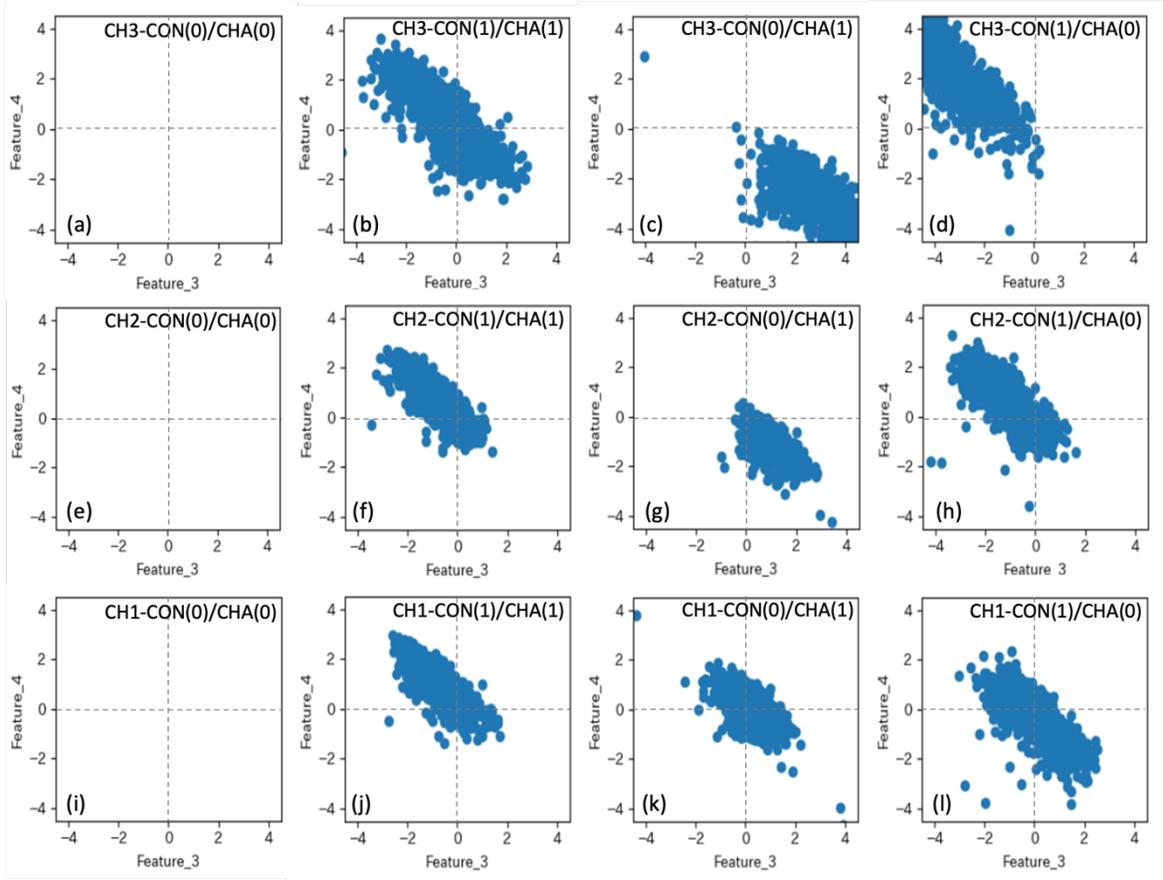

Fig. 9. Anomaly patterns in 2D feature space: (a-d) CH3 results for different operating conditions; (e-h) CH2 results; (i-l) CH1 results. Quantum kernel mapping reveals distinct quadrant-based clustering for different anomaly types.

The results for channel CH2 showed similar but less definitive patterns. While the normal/anomaly case produced results consistent with CH3, the anomaly/normal scenario resulted in anomaly points that, while primarily concentrated in the fourth quadrant, also showed some spread into the second quadrant. Furthermore, the abnormal/abnormal case closely resembled the abnormal/normal pattern, making clear differentiation between these two states more challenging for CH2. capability can be attributed to the spatial positioning of the microphone relative to the sound sources and the resulting differences in sound propagation and interference patterns. Specifically, CH2 appeared to provide worse directional discrimination between the two anomaly sources, likely due to less favorable acoustic positioning relative to the equipment, compared to CH3.

When observing only the CH1 case, comparing the three scenarios with CON/CHA=1/1, 0/1, and 1/0, all plots extend in directions toward the second and fourth quadrants, creating a state where separation is not possible. However, when considering the transition from CH3 to CH2 and then to CH1, plots with CON/CHA=1/0 tend to shift from the second quadrant to the fourth quadrant. Meanwhile, plots with CON/CHA=0/1 move from the fourth quadrant to the second quadrant, resulting in a state where they span across both quadrants in CH1.

These results demonstrate that not only can the quantum kernel approach successfully detect the presence of anomalies at greater distances than classical methods, but it can also differentiate between multiple types of anomalies based on their characteristic patterns in feature space: a capability that significantly enhances the practical value of such systems in complex manufacturing environments. Below, we will focus on CH3.

The distinct quadrant clustering of different anomaly types in feature space is a particularly valuable property of the quantum kernel approach. This spatial separation enables not only binary anomaly detection (normal vs. anomalous), but also multi-class classification of anomaly types without the need for explicit training of anomaly samples. This is a significant advantage in manufacturing environments where collecting comprehensive anomaly datasets is challenging.

The consistent appearance of CON anomalies in the second quadrant and CHA anomalies in the fourth quadrant suggests that the quantum kernel has identified fundamental differences in the acoustic signatures of these failure modes. This natural emergence of interpretable feature space organization is particularly noteworthy given that the model was trained only on normal data and had no explicit information about the different types of anomalies during training.

From a practical perspective, this quadrant-based classification provides maintenance personnel with an intuitive diagnostic tool. By simply observing which quadrant an anomaly point appears in, technicians can immediately identify which equipment requires attention without needing to manually investigate multiple devices. This capability could significantly reduce response times and maintenance costs in complex manufacturing environments with numerous interacting systems.

*F. Limitations and Impairment Analysis*

While the quantum kernel showed good performance, some conditions may lead to degradation. 1) Signal-to-Noise Ratio Threshold: Performance degradation was observed when background noise exceeded 85% of the total signal power (tested with additional white noise injection). 2) Feature Space Saturation: When tested with AR coefficient $p > 8$, the performance of the quantum kernel plateaued, suggesting limitations in the expressive power of quantum circuits. 3) Interference Patterns: In environments with five or more simultaneous sound sources, the selectivity of the directional microphone decreased, affecting both classical and quantum approaches. These results suggest that while the quantum kernel is more robust than classical approaches, it is not universally superior under all conditions.

## VI. Conclusion and outlook

To realize smart manufacturing using quantum machine learning, we aim to detect multiple anomalies using multiple non-contact sensors as shown in fig.1. In this study, we demonstrated the potential to individually detect two anomaly sounds using a single directional microphone (non-contact sensor) with an AR model incorporating a quantum kernel. In actual production environments, non-contact sensors are expected to be installed at distances exceeding 3m from equipment. In such cases, while anomaly detection using AR models with classical kernels is difficult, AR models with quantum kernels can potentially achieve high accuracy even when sensors are placed at a distance. We denoted that a single non-contact sensor could detect two anomalies. Each manufacturing device has the potential for multiple anomalies to occur. As shown in Fig.9, displaying anomaly occurrences on a two-dimensional plane makes detection easier for production site personnel.

While we have demonstrated the potential to individually detect two anomaly sounds, we aim to expand this work to detect many anomaly sounds in the future. Time series data is used in various fields such as medical data [25], [26], financial analysis [27], [28], and weather forecasting [29], [30], where anomaly detection is a critical challenge. We intend to utilize the rich expressive power of quantum kernel to solve various challenges in these domains. As fault-tolerant quantum computing (FTQC) capabilities mature, we anticipate the development of fully quantum computational methods that fundamentally transcend classical approaches by implementing end-to-end quantum algorithms for industrial anomaly detection.

Next, we will denote limitations and future directions. Several constraints in the scope of this study affect generalizability. First, tests with only two types of machines (conveyor and chain belt) do not provide sufficient evidence of broader applicability. Future studies should evaluate performance on a variety of manufacturing equipment with different acoustic characteristics. Second, relying solely on acoustic data may limit applicability in environments where acoustic-based detection is impractical. Third, the quadrant-based visualization approach may not scale to scenarios with more than four types of anomalies, requiring consideration of higher-dimensional feature space representations. Fourth, the controlled laboratory environment may not encompass all industrial environmental conditions.

Future work will focus on several promising directions. First, we plan to expand our approach to handle a greater diversity of anomaly types across more manufacturing equipment, testing the quadrant-based classification approach. Second, we will investigate the potential for deploying this system on near-term quantum hardware, assessing whether the theoretical advantages observed in simulation can be realized on actual quantum processors despite current hardware limitations. Finally, we aim to develop hybrid approaches that can leverage classical techniques with quantum kernel methods to optimize performance while minimizing quantum resource requirements.

As quantum computing hardware continues to advance, we anticipate that quantum-enhanced anomaly detection will become a cornerstone technology in next-generation manufacturing systems. The approach demonstrated in this paper represents an important first step toward realizing the full potential of quantum technologies in industrial settings, potentially bridging the gap between theoretical quantum advantages and practical manufacturing applications.

## Acknowledgment

This work was supported by the New Energy and Industrial Technology Development Organization (NEDO) [Grant No. JPNP23003] and by the Center of Innovations for Sustainable Quantum AI (JST) [Grant No. JPMJPF2221].

## Data and Code Availability

The datasets generated and analyzed during this study, including raw audio recordings and processed autoregressive coefficient features, are available from the corresponding author upon reasonable request for research purposes. The quantum kernel implementation code used for analysis will also be shared to ensure reproducibility. Requests should include a brief description of the intended use and institutional affiliation.